\documentclass[aps,prb,preprint,amsmath,amssymb,%
superscriptaddress,unsortedaddress,%
showpacs]{revtex4}

\usepackage{graphicx}

\newcommand{\STR}[3]{{#1}({#2}$\times${#3})}

\begin{document}

\title{The effect of Fe atoms on the adsorption of a W atom on W(100)
  surface}

\author{J.~Houze}
\author{Sungho~Kim}
\affiliation{
  Department of Physics and Astronomy, 
  Mississippi State University,
  Mississippi State, MS 39762, USA
}
\affiliation{%
  Center for Advanced Vehicular Systems, 
  Mississippi State University, 
  Mississippi State, MS 39762, USA
}
\author{Seong-Jin~Park}
\affiliation{%
  Center for Advanced Vehicular Systems, 
  Mississippi State University, 
  Mississippi State, MS 39762, USA
}
\author{Randall~M.~German}
\author{M.~F.~Horstemeyer}

\affiliation{
  Department of Mechanical Engineering,
  Mississippi State University,
  Mississippi State, MS 39762, USA
}
\affiliation{%
  Center for Advanced Vehicular Systems, 
  Mississippi State University, 
  Mississippi State, MS 39762, USA
}

\author{Seong-Gon~Kim}\email{kimsg@ccs.msstate.edu} 
\affiliation{
  Department of Physics and Astronomy, 
  Mississippi State University,
  Mississippi State, MS 39762, USA
}
\affiliation{%
  Center for Advanced Vehicular Systems, 
  Mississippi State University, 
  Mississippi State, MS 39762, USA
}

\date{\today}

\begin{abstract}
  We report a first-principles calculation that models the effect of
  iron (Fe) atoms on the adsorption of a tungsten (W) atom on W(100)
  surfaces. The adsorption of a W atom on a clean W(100) surface is
  compared with that of a W atom on a W(100) surface covered with a
  monolayer of Fe atoms.  The total energy of the system is computed
  as the function of the height of the W adatom.  Our result shows
  that the W atom first adsorbs on top of the Fe monolayer.  Then the W
  atom can replace one of the Fe atoms through a path with a moderate
  energy barrier and reduce its energy further.  This intermediate
  site makes the adsorption (and desorption) of W atoms a two-step
  process in the presence of Fe atoms and lowers the overall
  adsorption energy by nearly 2.4~eV.  The Fe atoms also provide a
  surface for W atoms to adsorb facilitating the diffusion of W atoms.
  The combination of these two effects result in a much more efficient
  desorption and diffusion of W atoms in the presence of Fe atoms.
  Our result provides a fundamental mechanism that can explain the
  activated sintering of tungsten by Fe atoms.
\end{abstract}


\maketitle

\section{Introduction}

The sintering of tungsten can be enhanced by the addition of small
amounts of alloying elements (0.5-1.0\%) from the iron group metals
such as Fe, Co, and Ni. This phenomenon is called ``activated
sintering.''\cite{Brophy:1961, German:1976} These additive species
lower the activation energy for sintering, resulting in a much lower
sintering temperature, shorter sintering time, or better
properties.\cite{German:2005} The classical theory explains this by
the enhancement of grain-boundary diffusion in tungsten due to the
presence of the respective element in the grain boundary.  High
densities (up to 99\% of bulk) can be obtained even at 1100~$^\circ$C
(at this temperature, tungsten compacts are commonly presintered).
\cite{Panichkina:1967, Samsonov:1969, German:2005} Because of high
melting point and large surface energy, tungsten has been considered
to be one of the best substrates to grow thin magnetic films on and
consequently its interaction with magnetic materials including Fe has
been studied extensively.\cite{Dennler:2005, Spisak:2004}

In this study, we investigated the activated sintering of tungsten by
iron from a quantum mechanics point of view.  Adsortion and desorption are
two of the main mechanisms of sintering. The aim of the present work
is to explore the morphology and energetics of thin Fe layers on W
surfaces and their effect on the adsorption/desorption of W atoms and
eventually on the sintering of tungsten.  Our approach is based on
density functional theory and we show that our atomistic
first-principles modeling and simulation can supplement the
experiments and provide a fundamental mechanism of the activated
sintering of tungsten by iron.  Our present study is focused on the
adsorption of tungsten on W(100) surfaces.  The adsorption of W atoms
on a clean W(100) surface is compared with the one on the same surface
covered with a monolayer of Fe atoms.  The surface morphology and
energy barriers were calculated and analyzed to elucidate the role of
Fe atoms in the adsorption of W atoms.

\section{Methods}
\label{sec:Methods}

The adsorption energy of a single adatom $E_{\text{ads}}$
at height $z$ is given by
\begin{equation}
  \label{eq:E_ads}
  E_{\text{ads}}(z) = E_{\text{tot}}(z) - E_{\text{tot}}(\infty)
\end{equation}
where $E_{\text{tot}}(z)$ is the total energy of the structure with
the adatom adsorbed at height $z$ on the surface and
$E_{\text{tot}}(\infty)$ is the total energy of the same surface with
the adatom at an infinite distance.

All \textit{Ab initio} total-energy calculations and geometry
optimizations are performed within density functional theory (DFT)
using Bl\"{o}chl's all-electron projector augmented wave (PAW)
method\cite{Blochl:1994:PAW} as implemented by Kresse et
al.\cite{Kresse:1999:PhysRevB.59.1758} For the treatment of electron
exchange and correlation, we use the generalized gradient
approximation (GGA) of Perdew et al.\cite{Perdew:1996} The Kohn-Sham
equations are solved using a preconditioned band-by-band
conjugate-gradient (CG) minimization.\cite{Kresse:1993} The plane-wave
cutoff energy is set to at least 350~eV in all calculations.  The
tungsten lattice constant of 3.175 \AA\ we obtained is slightly larger
than the experimental value of 3.165 \AA.\cite{AIP-Handbook} We use a
standard supercell technique, modeling the W(100) surface with
\STR{}{3}{3} surface unit cell by a slab consisting of seven substrate
layers, separated by 10.0 \AA\ of vacuum.  Atoms in the bottom layer
are fixed at their bulk positions, while all other atoms are allowed
to relax until the root-mean-square (rms) force is less than
0.001~eV/\AA.  The Brillouin zone is sampled with a density equivalent
to at least 81 $k$-points in \STR{}{1}{1} surface Brillouin zone using
the Monkhorst-Pack scheme.\cite{Monkhorst:1976} A Fermi-level smearing
of 0.2~eV was applied using the Methfessel-Paxton
method.\cite{Methfessel:1989:PhysRevB.40.3616}

\section{Results and Discussion}
\label{sec:Results}

\begin{figure}[!tbp]
  \includegraphics[width=.64\columnwidth]{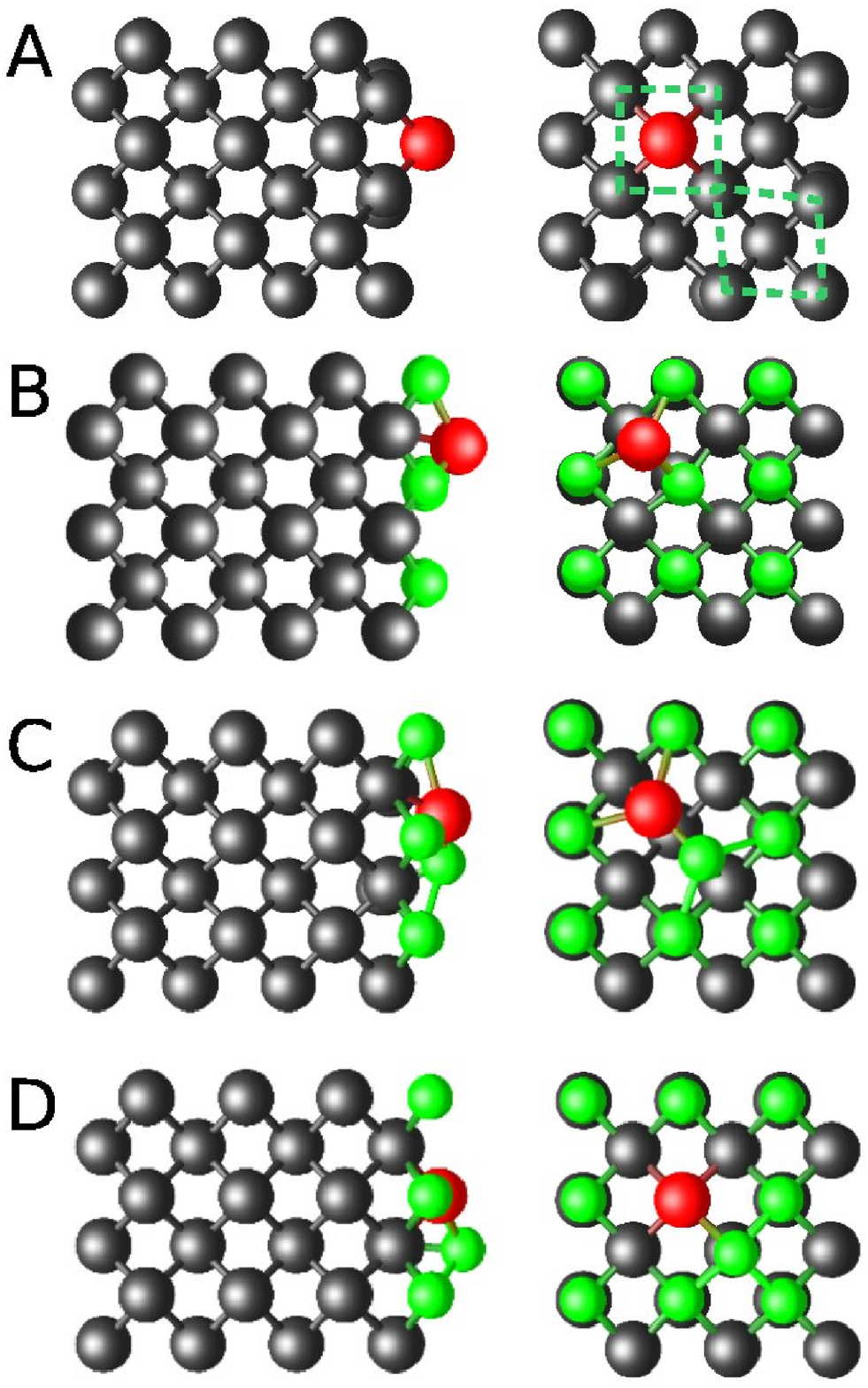}
  \caption{\label{fig:W-adsorption} (color online) Adsorption of W on
    (A) clean W(100) and (B-D) Fe monolayer on W(100) surfaces.  The
    side views (left column) and top views (right column) are given.
    \STR{}{3}{3} surface unit cells are used for our calculations as
    shown in top views.  Black spheres represent W atoms in the
    substrate layers while a red sphere represents the W adatom.  Fe
    atoms are represented by smaller green spheres.  (A) W adatom is
    adsorbed on a clean W(100) surface. (B) W adatom is adsorbed on a
    Fe monolayer covering W(100) surface. (C) W adatom is pushed in
    toward the closest Fe atom, which is displaced from its place.
    The W and Fe atoms form a dimer that is parallel to W(100)
    surface. (D) W adatom completely replaced the Fe atom and the
    displaced Fe atom is now adsorbed on the Fe monolayer.  Two
    deltoids in (A) show the distortion of nearest neighbor W atoms
    bonds due to surface reconstruction.}
\end{figure}

In this study, we focus on the adsorption of W atoms on W(100) and
Fe/W(100) surfaces.  Fig.~\ref{fig:W-adsorption} shows the optimized
structures of adsorbed W atoms in different configurations. In
Fig.~\ref{fig:W-adsorption}(A), a W adatom is adsorbed on a clean
W(100) surface, while Fig.~\ref{fig:W-adsorption}(B-D) show three
different configurations of a W adatom adsorbed on W(100) surface
covered by a Fe monolayer.  For each configuration, all atoms are
fully relaxed except the atoms in the bottom layer.

W(100) surface is known to exhibit a
\STR{}{$\sqrt{2}$}{$\sqrt{2}$}$R$45$^{\circ}$ reconstruction at low
temperatures.\cite{Altman:PhysRevB.38.5211} It also has been observed
that the surface reconstruction vanishes locally upon the adsorption
of a Mn atom on W(100) surface.\cite{Spisak:2004} Our result is in
good agreement with these observations. Two deltoids in
Fig.~\ref{fig:W-adsorption}(A) show the distortion of nearest neighbor
bond angles due to surface reconstruction.  The upper-left deltoid
containing the W adatom is indistinguishable from a square reflecting
the fact that the reconstruction is suppressed due to the adsorption
of the W atom.  On the contrary, the lower-right deltoid has two equal
obtuse angles of 96.1$^{\circ}$ indicating the distortion of bond
angles due to surface reconstruction. However, because the deltoid is
adjacent to the W adatom adsorption site, the angle is smaller than
106$^{\circ}$ observed for clean W(100) surfaces.\cite{Dennler:2005}

\begin{figure}[!tbp]
  \includegraphics[width=.98\columnwidth]{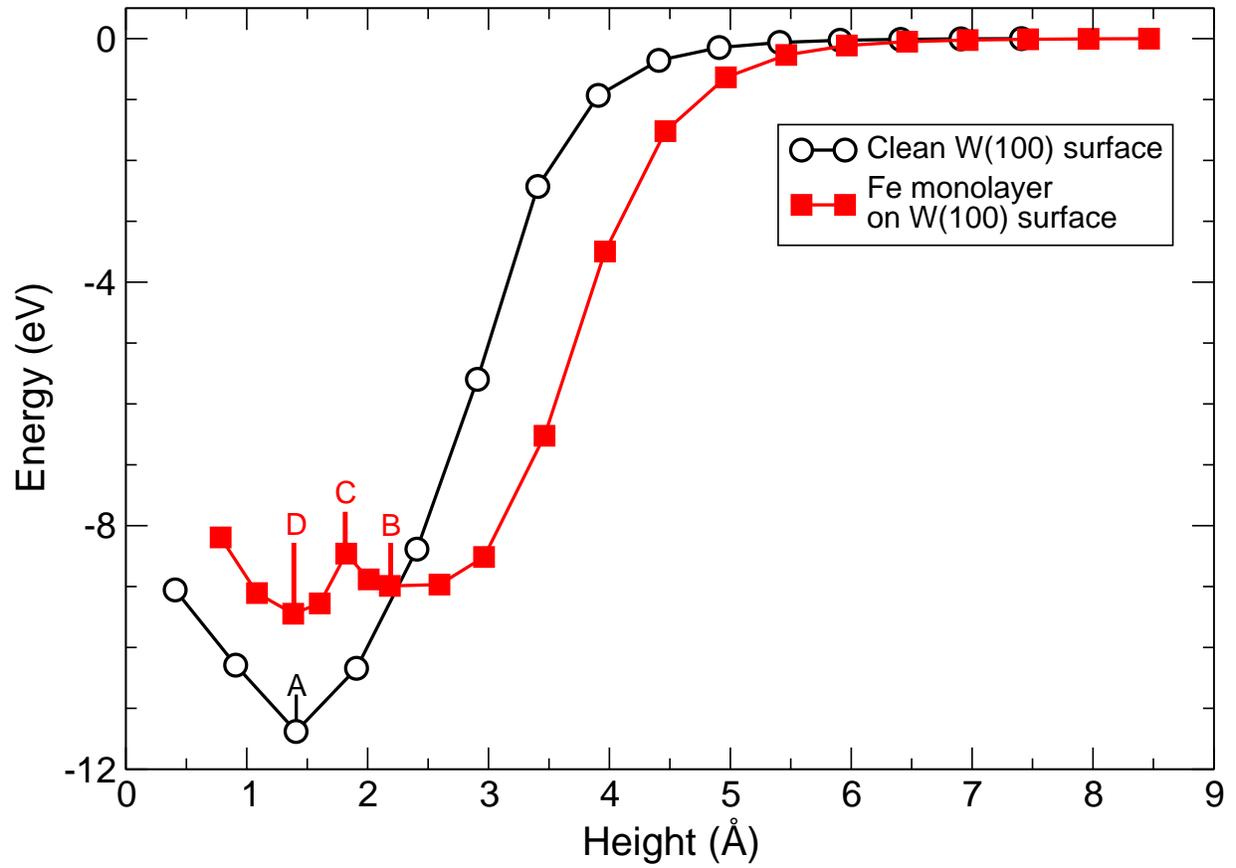}
  \caption{\label{fig:W-adsorption-energy} (color online) The
    adsorption energy $E_{\text{ads}}(z)$ of a W adatom on a clean W(100)
    surface (black open circles) and on a Fe monolayer on W(100)
    surface (red filled squares).  The height of W adatom $z$ is
    measured from the W atoms in the top layer of W(100) surface.  }
\end{figure}

The adsorption energies and the optimized height for these
configurations are listed in Table~\ref{tab:E-ads}.
Fig.~\ref{fig:W-adsorption-energy} also shows the adsorption energy of
a W adatom on two different surfaces, a clean W(100) surface and an Fe
monolayer on W(100) surface, as a function of the height.  For the
configurations (A-D), only the bottom layers are fixed and all other
atom positions are fully relaxed.  The other points in
Fig.~\ref{fig:W-adsorption-energy} are generated by additionally
fixing the height of the W adatom.

Fig.~\ref{fig:W-adsorption}(B) shows the morphology of a W atom
adsorbed on an Fe monolayer on the W(100) surface.  It turned out that
the most energetically favorable adsorption site for a W adatom is not
directly above the center of the square formed by four neighboring Fe
atoms, but closer to one of the surrounding Fe atoms.  In
Fig.~\ref{fig:W-adsorption}(C), the W adatom is pushed in toward the
closest Fe atom. Consequently, the Fe atom is displaced from its place
and forms a horizontal dimer with the W adatom.  Although
configuration $C$ has a higher adsorption energy than that of
configuration $B$ by $E_{\text{C}} - E_{\text{B}} = 0.53$~eV as shown
in Table~\ref{tab:E-ads}, configuration $C$ is stable against small
perturbation of atomic positions of the W-Fe dimer.  When the W adatom
is pushed in further, it replaces the Fe atom completely and the
displaced Fe atom will adsorb above the Fe monolayer, as shown as
configuration $D$ in Fig.~\ref{fig:W-adsorption}(D).

Our result is consistent with the experimental observation that the
addition of a small amount of Fe atoms enhances the sintering of
tungsten.  Our study shows that there are two main effects of Fe atoms
on tungsten sintering.  First, the existence of a stable adsorption
site on an Fe monolayer (configuration $B$) leads to more efficient
desorption of W atoms.  This means that the desorption of W atoms
occurs in two stages when Fe atoms are present.  Starting from
configuration $D$, W atoms first move to configuration $B$ by passing
through the Fe layers via the pathway $D \to C \to B$.  The W atoms
overcome a moderate size energy barrier of $E_{\text{C}} -
E_{\text{D}} = 0.99$~eV utilizing the thermal excitation readily
available during sintering process.  At the second stage, W atoms can
be desorbed from configuration $B$.  Compared to the direct desorption
from configuration $A$ in the absence of Fe atoms, this multi-stage
desorption through configuration $B$ is much more efficient because
the adsorption energy of W atom at $B$ is significantly lower than at
$A$: $E_{\text{B}} - E_{\text{A}} = (E_{\text{B}} - E_{\text{D}}) +
(E_{\text{D}} - E_{\text{A}}) = 0.46 + 1.93 = 2.39$~eV (See
Table~\ref{tab:E-ads}).  Second, the existence of stable adsorption
sites on Fe monolayer makes the diffusion of W atoms on Fe atoms
available for an efficient sintering process.  The smaller adsorption
energy of a W atom on an Fe monolayer strongly suggests that the Fe
thin film layer will provide a much smoother platform for the adsorbed
W atoms to diffuse.  This enhanced diffusion will facilitate the
movement of W atoms from one place to another.  A further detailed
investigation will be required to elucidate the effect of Fe atoms on
diffusion of W atoms and it will be reported elsewhere.  Comparison
with different activation agents, such as Co or Ni, will also provide
valuable information in validating the mechanism proposed from the
present work.

\section{Conclusions}
\label{sec:Conclusion}

We have presented a first-principles DFT investigation of the
structure and energetics of a W adatom on W(100) and Fe/W(100)
surfaces.  We found that W atoms adsorbs on an Fe monolayer. Fe atoms
also reduce the adsorption energy of W adatoms on W(100) surface
substantially.  Fe thin film layers also allow W atoms to pass through
easily and thus facilitate the diffusion of W atoms.

\section{Acknowledgment}

The authors are grateful to the Center for Advanced Vehicular Systems
at Mississippi State University for supporting this study.  Computer
time allocation has been provided by the High Performance Computing
Collaboratory (HPC$^2$) at Mississippi State University.  


\newpage

\begin{table}[!tbp]
  \caption{\label{tab:E-ads} The adsorption energy $E_{\text{ads}}$
    and the optimized height $z$ for different configurations  
    in Fig.~\ref{fig:W-adsorption}.  Energy differences between
    different configurations are also listed.}
  \begin{ruledtabular}
    \begin{tabular}{crrc}
      Config. & $E_{\text{ads}}$ (eV) & $z$ (\AA) & $\Delta E$ (eV) \\ 
      \hline
      A & -11.38 & 1.41 & $E_{\text{D}} - E_{\text{A}} = 1.93$ \\
      D &  -9.45 & 1.38 & $E_{\text{C}} - E_{\text{D}} = 0.99$ \\
      C &  -8.46 & 1.82 & $E_{\text{B}} - E_{\text{C}} = -0.53$ \\
      B &  -8.99 & 2.18 & $E_{\text{B}} - E_{\text{D}} = 0.46$ \\
    \end{tabular}
  \end{ruledtabular}
\end{table}

\end{document}